\documentclass[aps,pra,twocolumn,superscriptaddress,showpacs,amsmath,amssymb]{revtex4}
\usepackage[usenames,dvipsnames]{color}
\usepackage{epsf}
\usepackage{bm}
\usepackage{dcolumn}
\usepackage{latexsym}
\usepackage{amsmath}
\usepackage{amsfonts}
\usepackage{amssymb}
\usepackage{graphicx}
\usepackage[active]{srcltx}
\usepackage[autostyle]{csquotes}
\usepackage{braket}
\usepackage{textcomp}
\newcommand{\be}{\begin{eqnarray}}
\newcommand{\ee}{\end{eqnarray}}

\definecolor{darkred}{rgb}{.8,0,0}

\definecolor{darkblue}{rgb}{0,0,.7}

\begin{document}
%
\title{Making the most of time in quantum metrology: concurrent state preparation and sensing}
%
\author{Anthony J. Hayes}
\email{Anthony.Hayes@sussex.ac.uk}
\affiliation{Department of Physics and Astronomy, University of Sussex, Brighton, BN1 9QH, United Kingdom\\}
\affiliation{NTT,Basic Research Laboratories, 3-1 Morinosato-Wakamiya, Atsugi, Kanagawa 243-0198, Japan}
\author{Shane Dooley}
\email{dooleysh@gmail.com}
\affiliation{National Institute of Informatics, 2-1-2 Hitotsubashi, Chiyoda-ku, Tokyo 101-8430, Japan}
\author{William J. Munro$^{\:2,\:3}$}

\author{Kae Nemoto$^{\:3}$}

\author{Jacob Dunningham$^{\:1}$}

\begin{abstract}
A quantum metrology protocol for parameter estimation is typically comprised of three stages: probe state preparation, sensing and then readout, where the time required for the first and last stages is usually neglected. In the present work we consider non-negligible state preparation and readout times, and the tradeoffs in sensitivity that come when a limited time resource $\tau$ must be divided between the three stages. To investigate this, we focus on the problem of magnetic field sensing with spins in one-axis twisted or two-axis twisted states. We find that (accounting for the time necessary to prepare a twisted state) by including entanglement, which is introduced via the twisting, no advantage is gained unless the time $\tau$ is sufficiently long or the twisting sufficiently strong. However, we also find that the limited time resource is used more effectively if we allow the twisting and the magnetic field to be applied concurrently which is representative of a more realistic sensing scenario. We extend this result into the optical regime by utilizing the exact correspondence between a spin system and a bosonic field mode as given by the Holstein-Primakoff transformation.
\end{abstract} 
\vspace{-1mm}
\maketitle
%
%
%
\section{Introduction~\label{sec1}}
%
Quantum metrology utilises non-classical effects in order to enhance the precision to which measurements can be made \cite{giovannetti2006quantum}. This has had many useful applications in fields as diverse as gravitational wave detection \cite{schnabel2010quantum,caves1981quantum,abbott2016observation}, magnetometry \cite{mussel2014scalable} and biological sensing \cite{wolfgramm2013entanglement,crespi2012measuring,taylor2013biological,barry2016optical}. If quantum metrology is to become a widespread technology, the theoretical models should incorporate further, more realistic aspects of the system. In this paper, we consider non-negligible state preparation and readout times, and we investigate the tradeoffs in sensitivity when a limited time resource must be divided between the various stages of a quantum metrology protocol.

A quantum metrology protocol is typically ordered into three stages: $i)$ Probe state preparation, in which quantum mechanical correlations are introduced to a system that will be used as a probe. Examples include the generation of spin squeezed states \cite{dooley2016hybrid} or of cat states \cite{dooley2013collapse,munro2002weak}. $ii)$ Sensing, in which the probe is subject to, and consequently altered by, a parameter of interest. The quantum mechanical correlations introduced in the preparation stage increase the probe's susceptibility to alterations caused by this parameter beyond classical limits. $iii)$ Readout, in which a final measurement is made on the altered probe state enabling estimation of the parameter of interest. 

The three stages of the protocol take a combined time $\tau$. Usually, the state preparation and readout times are assumed to be negligible, so that the total time $\tau$ can be devoted to the sensing stage. If the state preparation and readout times are non-negligible, however, $\tau$ should be divided between the three stages \cite{dooley2016quantum}. This leads to a trade-off since, for example, too much time given to state preparation subtracts from the available time for sensing, while too little time given to state preparation may not allow enough time to generate the most sensitive state.

\begin{figure}
	\includegraphics[width=1\columnwidth]{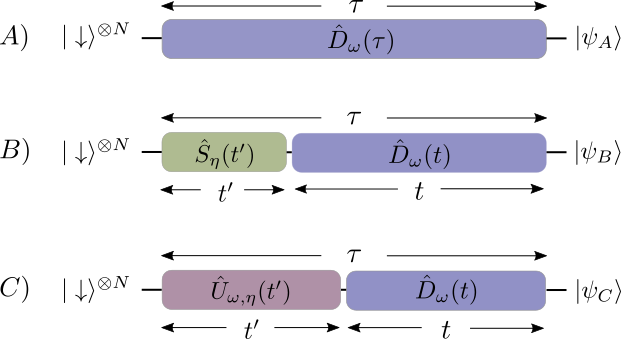}
	\caption{In scheme $A$, the magnetic field is applied over the entire time $\tau$, by the operation $\hat{D}_\omega (\tau)$. The spins remain in a separable state throughout. In scheme $B$, the two-axis twisting operation $\hat{S}_\eta (t')$ generates a sensitive entangled state before exposure to the magnetic field through $\hat{D}_\omega (t)$. In scheme $C$ the spins are subject to the operation $\hat{U}_{\omega,\eta}(t')$ (as defined in section \ref{II.C}) which exposes them to the magnetic field during the twisting operation. Each scheme ends with a measurement of the final state $\ket{\psi_{i}}$ ($i\in\{A,B,C\}$), which we assume can be done in a negligible time. For a fair comparison, between the three schemes, each is constrained by the time $\tau$.}
	\label{fig:schemes_TAT}
\end{figure}

In this paper we explore this problem in the context of magnetic field sensing with a probe consisting of $N$ spin-1/2 particles. We compare three different strategies depicted in Fig.\ref{fig:schemes_TAT}. In scheme $A$, the magnetic field is probed with a separable state of the spins, which we assume can be prepared and read-out in a negligible time. In scheme $B$, a non-negligible preparation time is used to generate a twisted (i.e., entangled) spin state \cite{Kit-93}, before exposing it to the magnetic field. Finally, in scheme $C$, we investigate whether the limited time resource $\tau$ can be used more effectively by allowing the twisting operation and the magnetic field to be applied simultaneously. By a combination of numerical and analytical results, we find that scheme $C$ is indeed a more effective use of the limited time resource than scheme $B$. Comparing schemes $B$ and $C$ to scheme $A$, we also find that --- taking the non-negligible state preparation times into account --- twisting gives no improvement in sensitivity unless the total time resource $\tau$ is sufficiently long, or the twisting sufficiently strong. In section \ref{sec:TAT} we consider schemes where the entanglement is generated by two-axis twisting and the final readout is optimised over all possible measurements. In section \ref{sec:OAT}, motivated by the recent work of Davis and co-workers \cite{davis2016approaching}, we consider an arguably more realistic scheme where the entanglement is generated by one-axis twisting and the readout is by a so-called echo measurement. Conclusions are given in section \ref{sec8}.  

\section{Magnetic field sensing and Two-Axis Twisting \label{sec:TAT}}

In this section we consider our schemes $A$, $B$ and $C$, illustrated in Fig. \ref{fig:schemes_TAT}. Before describing each scheme in detail, it is useful to introduce the collective spin operators $\hat{J}_{\mu}=\sum_{i=1}^{N}\hat{\sigma}_{i}^{(\mu)}$, where $\sigma^{(\mu)}_{i}$ are the Pauli spin operators for the $i$'th spin-1/2 particle with $\mu\in\{x,y,z\}$. Eigenstates of the $\hat{\sigma}^{(z)}$ operator are denoted $\ket{\uparrow}$ and $\ket{\downarrow}$. Furthermore, we can define the raising and lowering operators $\hat{J}_{\pm}=\hat{J}_{x}\pm i \hat{J}_{y}$. As shown in Fig. \ref{fig:schemes_TAT}, in all three schemes we assume that the initial ``unprepared'' probe state is the coherent spin state $\ket{\downarrow}^{\otimes N}$ and that the final state is $\ket{\psi_j}$ ($j\in\{A,B,C\}$). For simplicity, in this section we assume that the final readout of the state $\ket{\psi_j}$ takes a negligible amount of time.

To quantify the magnetic field sensitivity of the scheme $j\in\{A,B,C\}$, we make use of the quantum Cramer-Rao inequality \cite{paris2009quantum,wiseman2009quantum} $\delta \omega_j \geq 1/\sqrt{\nu F_j}$, where we have used $\omega$ to denote the \textit{scaled} magnetic field; the frequency $\omega = \gamma B$ is proportional to the magnetic field $B$, so that the problem of estimating $\omega$ is the same as the problem of estimating $B$ when the gyromagnetic ratio $\gamma$ is known. This gives an upper bound on the error $\delta\omega_j$ of the estimate of the scaled magnetic field $\omega$. The Cramer-Rao bound holds for sufficiently large number of of repeats of the measurement scheme $\nu$. The quantity $F_j$ is the quantum Fisher information, which around $\omega \approx 0$ is given by: \begin{equation} F_j = 4 \left[ \langle \partial_\omega\psi_j | \partial_\omega \psi_j \rangle + |\langle\psi_j|\partial_\omega\psi_j\rangle|^2 \right]_{\omega = 0} \label{eq:QFI} \end{equation} where $|\partial_\omega\psi_j\rangle=\frac{\partial}{\partial\omega}\ket{\psi_{j}}$. We can quantify the sensitivity by the dimensionless quantity \begin{equation} (\sqrt{\nu}\tau\delta\omega_j)^{-1} \leq \sqrt{F_j}/\tau , \label{eq:sens_upper_bound} \end{equation} where the upper bound follows from the quantum Cramer-Rao inequality. Eq. \ref{eq:sens_upper_bound} is valid when $\nu\gg1$ and we note that if the final measurement of the state $\ket{\psi_j}$ is optimised, it is possible to saturate the inequality.


We now describe schemes $A$, $B$ and $C$ in detail, and calculate the dimensionless sensitivity Eq. \ref{eq:sens_upper_bound} in each case. 


\subsection*{Scheme $A$}

In scheme $A$, the initial state $\ket{\downarrow}^{\otimes N}$ evolves by a scaled magnetic field $\omega$ (in the $y$-direction) for the total time $\tau$, giving the final state: \begin{equation} \ket{\psi_A} = \hat{D}_\omega (\tau) \ket{\downarrow}^{\otimes N} , \label{eq:psi_f_A} \end{equation} where $\hat{D}_\omega (\tau) \equiv \exp[ -i\tau\hat{H}_{\omega}/\hbar]$ and $\hat{H}_\omega = \hbar\omega\hat{J}_y /\sqrt{N}$. (Note that for later convenience the Hamiltonian $\hat{H}_\omega$ has been scaled by a factor of $1/\gamma\sqrt{N}$). The unitary $\hat{D}_\omega (\tau)$ causes a rotation of the ``unprepared'' state around the $y$-axis by an angle $\phi=\omega\tau/\sqrt{N}$, where $\omega$ is to be estimated. Clearly there is no entanglement between spins at any time in this scheme. Calculating the quantum Fisher information by Eq. \ref{eq:QFI} gives: \begin{equation} (\sqrt{\nu}\tau\delta\omega_{A})^{-1} \leq \sqrt{F_{A}}/\tau = 1 . \label{eq:sens_A_TAT} \end{equation} This is the benchmark against which we compare the sensitivities of schemes $B$ and $C$.





\subsection*{Scheme $B$}

\begin{figure*}
	\includegraphics[width=2\columnwidth]{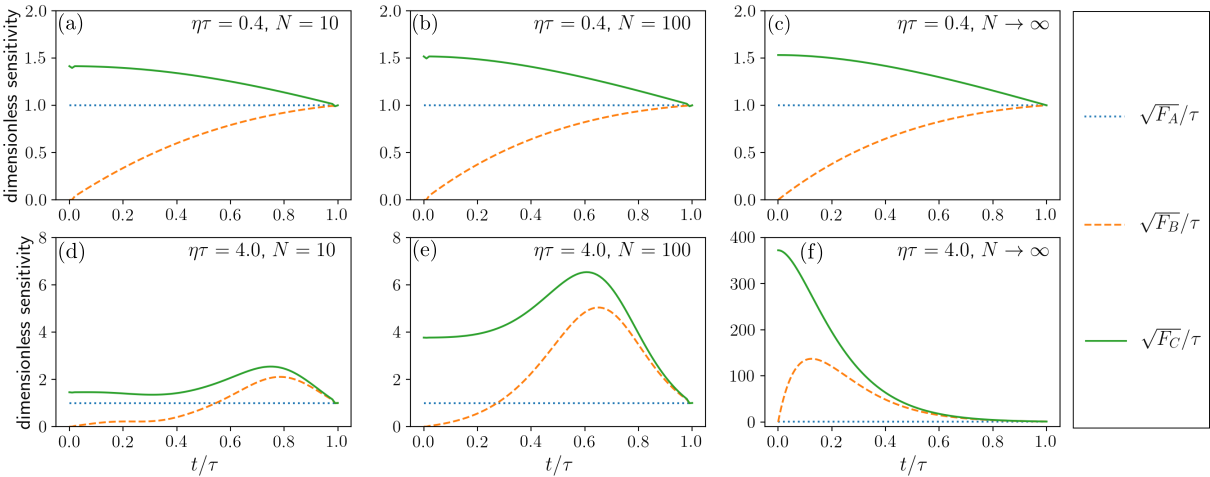}
	\caption{These plots show that for a sufficiently small value of $\eta\tau$ (e.g. $\eta\tau = 0.4$ in the upper plots), scheme $B$ gives no improvement over scheme $A$. For a sufficiently large value of $\eta\tau$ (e.g. $\eta\tau = 4$ in the lower plots), both scheme $B$ and scheme $C$ can give a better sensitivity than scheme $A$ (i.e., the two-axis twisting state preparation is worthwhile), if the sensing time $t/\tau$ is optimised.}
	\label{fig:sens_vs_t_TAT}
\end{figure*}

One of the main results in the field of quantum metrology is that we can, in principle, improve on scheme $A$ by generating an entangled state of the probe before exposing it to the magnetic field during the sensing period. When the entangled state preparation and readout times can be neglected, this is known to give a large improvement in the estimate of $\omega$ compared to scheme $A$. However, the extra time cost of preparing the entangled state is usually not taken into account. In scheme $B$ we include the time required for state preparation.

One class of entangled states are two-axis twisted (TAT) states \cite{ma2011quantum, Kit-93}. In our scheme $B$, starting from the initial state $\ket{\downarrow}^{\otimes N}$, the spins evolve by the TAT operation $\hat{S}_\eta(t') = \exp[-it' \hat{H}_{\eta}/\hbar]$ for a state preparation time of duration $t'$. Here $\hat{H}_\eta = i\hbar\eta (\hat{J}_{-}^2 - \hat{J}_{+}^2)/N$ is the two-axis twisting Hamiltonian, which has been scaled by a factor of $1/N$ for later convenience, and $\eta$ is the twisting strength. For small $\eta t'$, this operation generates squeezed states with a reduced standard deviation of the spin observable $\hat{J}_x$ \cite{ma2011quantum, Kit-93}. Such states are highly sensitive to spin rotations around the $y$-axis, since only a small rotation is necessary to result in a state that is easily distinguishable from the state prior to the small rotation. For larger values of $\eta t'$, two-axis twisting generates ``over-squeezed'' states, including Schr\"{o}dinger cat states. Over-squeezed states are also highly-sensitive to spin rotations around the $y$-axis and, if state preparation and readout times are neglected, can give sensitivity of the scaled magnetic field at the Heisenberg limit $(\sqrt{\nu}\tau\delta\omega)^{-1} = \sqrt{N}$ which has been scaled here by a factor of $1/\sqrt{N}$ due to the prior scaling introduced in the Hamiltonians. 

%
%

After the spins are prepared in the two-axis twisted state, they are exposed to the magnetic field for a time $t$, resulting in a rotation of the state around the spin $y$-axis by $\hat{D}_\omega (t)=\exp [ - i t\hat{H}_{\omega}/\hbar ]$. The final state is thus: \begin{equation} \ket{\psi_B} = \hat{D}_\omega(t) \hat{S}_\eta(t') \ket{\downarrow}^{\otimes N} . \end{equation} To ensure that the total time of scheme $B$ is limited to $\tau$, we have $t' = \tau - t$. We note that if $t = \tau$, there is no two-axis twisting and scheme $B$ reduces to scheme $A$.

Since an exact analytic expression for the quantum Fisher information $F_B$ is unknown, we calculate it numerically. An examination of the parameters of scheme $B$ shows that the dynamics are completely determined by only three independent, dimensionless variables: $N$ (the number of spins), $t/\tau$ (the fraction of the total measurement time given to the sensing stage), and $\eta\tau$ (the total measurement time $\tau$ in units of $1/\eta$). We now explore the sensitivity in this parameter space. In Fig. \ref{fig:sens_vs_t_TAT}, the dashed oragne lines show $\sqrt{F_B}/\tau$ as a function of the sensing time $t/\tau$ for various choices of $\eta\tau$ and $N$. We notice that there are some values of $\eta\tau$ and $N$ for which scheme $B$ gives no advantage over scheme $A$ for any choice of sensing time $t/\tau$ [see Figs. \ref{fig:sens_vs_t_TAT}(a), \ref{fig:sens_vs_t_TAT}(b), and \ref{fig:sens_vs_t_TAT}(c)]. In these cases, the sensitivity of scheme $B$ approaches that of scheme $A$ only as $t/\tau \to 1$ (i.e., as scheme $B$ approaches scheme $A$). This shows that two-axis twisting does not always give improvements in sensitivity, when a non-negligible state preparation time is taken into account. However, for other values of $\eta\tau$ and $N$, it is clear that scheme $B$ does give improvements over scheme $A$, if the sensing time $t/\tau$ is carefully chosen [see Figs. \ref{fig:sens_vs_t_TAT}(d), \ref{fig:sens_vs_t_TAT}(e), and \ref{fig:sens_vs_t_TAT}(f)].

\begin{figure*}
	\includegraphics[width=2\columnwidth]{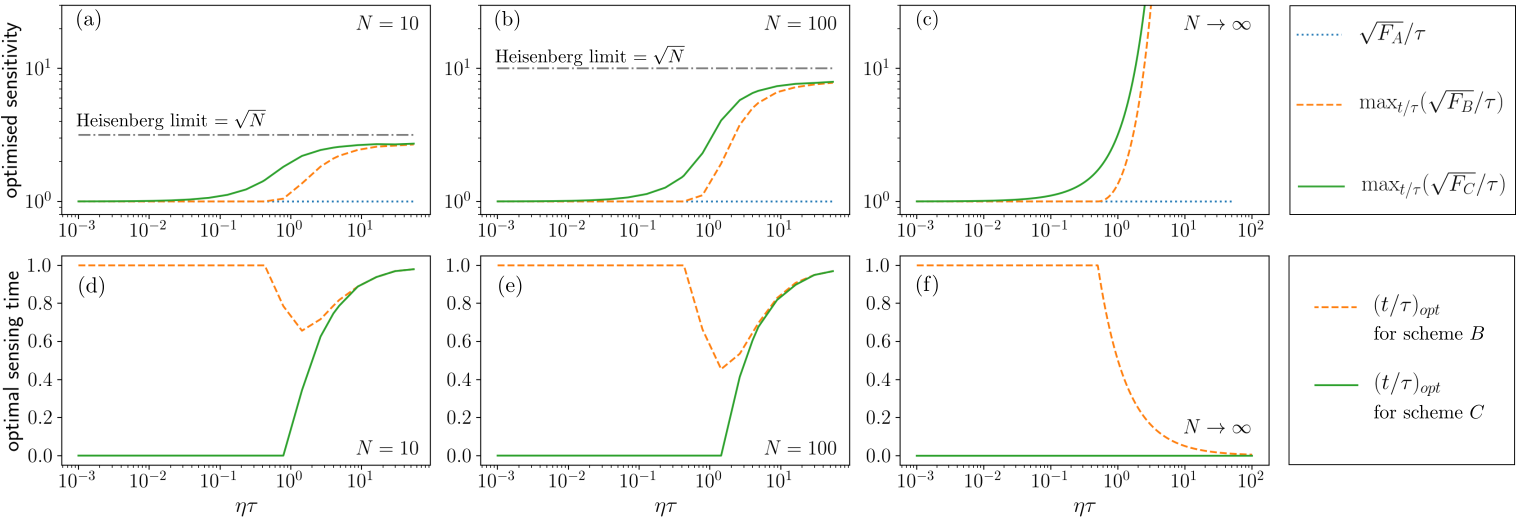}
	\caption{The upper plots show the optimised sensitivity $\max_{t/\tau}(\sqrt{F_i}/\tau)$ as a function of $\eta\tau$, and the lower plots show the corresponding optimal sensing times, $(t/\tau)_\text{opt}$. Comparison of schemes reveals that scheme $B$ gives no advantage over scheme $A$ for $\eta\tau\lesssim 0.5$. Scheme $C$, however, does better than scheme $A$ for all values of $\eta\tau$, although the advantage vanishes as $\eta\tau \to 0$.}
	\label{fig:opt_sens_vs_tau_TAT}
\end{figure*}

We can reduce the size of the parameter space and simplify the analysis by optimising over the sensing time $t/\tau$ for each value of $\eta\tau$ and $N$. This optimisation is done numerically and the results are plotted against $\eta\tau$ in Figs. \ref{fig:opt_sens_vs_tau_TAT}(a) and \ref{fig:opt_sens_vs_tau_TAT}(b), with the corresponding optimal sensing times $(t/\tau)_\text{opt}$ plotted in Figs. \ref{fig:opt_sens_vs_tau_TAT}(d) and \ref{fig:opt_sens_vs_tau_TAT}(e), respectively. These plots show that scheme $B$ gives no advantage over scheme $A$ if $\eta\tau \lesssim 0.5$, i.e., if the sensing time $\tau$ is sufficiently short or the twisting strength $\eta$ sufficiently weak. This conclusion follows from the observation that for $\eta\tau \lesssim 0.5$, the optimal sensing time is $(t/\tau)_\text{opt} = 1$, i.e., the full time $\tau$ is devoted to sensing, there is no two-axis twisting, and scheme $B$ reduces to scheme $A$.

If $\eta\tau\to\infty$ (the measurement time is infinitely long or the twisting is infinitely strong), any squeezed or over-squeezed state can be prepared in a negligible fraction of the total available time $\tau$. Indeed, Figs. \ref{fig:opt_sens_vs_tau_TAT}(a) and \ref{fig:opt_sens_vs_tau_TAT}(b) show that for $\eta\tau \gg 1$ the sensitivity approaches the Heisenberg limit, while Figs. \ref{fig:opt_sens_vs_tau_TAT}(d) and \ref{fig:opt_sens_vs_tau_TAT}(e) show that the state preparation time becomes a small fraction of $\tau$ (since the optimal sensing time $(t/\tau)_\text{opt}$ is close to, but not equal to, unity).


Although the analytic calculation of the quantum Fisher information $\sqrt{F_B}/\tau$ is intractable for arbitrary $N$, it is possible to calculate it in the limit $N\to\infty$. We find (see Appendix for details) that: \begin{eqnarray}  \sqrt{F_B}/\tau \stackrel{N\to\infty}{\longrightarrow} \frac{t}{\tau} e^{2\eta\tau (1 - t/\tau)} . \label{eq:F_B_N_inf} \end{eqnarray} Optimising Eq. \ref{eq:F_B_N_inf} over the sensing time $t/\tau$ gives different answers depending on whether $\eta\tau > 0.5$ or $\eta\tau \leq 0.5$. If $\eta\tau > 0.5$ we have: \begin{eqnarray} && \max_{t/\tau}(\sqrt{F_{B}}/\tau) \stackrel{N\to\infty}{\longrightarrow} \frac{e^{2\eta\tau - 1}}{2\eta\tau} , \\ && (t/\tau)_\text{opt} \stackrel{N\to\infty}{\longrightarrow} \frac{1}{2\eta\tau} . \end{eqnarray} If, however, $\eta\tau \leq 0.5$ we have \begin{eqnarray} && \max_{t/\tau}(\sqrt{F_{B}}/\tau) \stackrel{N\to\infty}{\longrightarrow} 1 , \\ && (t/\tau)_\text{opt}  \stackrel{N\to\infty}{\longrightarrow} 1 , \end{eqnarray} These quantities are plotted in Figs. \ref{fig:opt_sens_vs_tau_TAT}(c) and \ref{fig:opt_sens_vs_tau_TAT}(f). Comparison with the sensitivity $\sqrt{F_{A}}/\tau = 1$ for scheme $A$ shows that, in the $N\to\infty$ limit, preparation of a squeezed state via scheme $B$ gives an enhanced sensitivity only if $\eta\tau > 0.5$. If $\eta\tau \leq 0.5$, however, we have $(t/\tau)_\text{opt}  \stackrel{N\to\infty}{\longrightarrow} 1$ and the whole of the available time $\tau$ should be used for sensing without any squeezing (i.e., scheme $B$ reduces to scheme $A$), in broad agreement with the numerical results for finite $N$.

\subsection*{Scheme $C$}
\label{II.C}
During the state preparation stage in scheme $B$, the probe is not exposed to the magnetic field. This begs the question: can the limited time resource $\tau$ be used more efficiently by applying the magnetic field during the spin squeezing operation? This motivates our scheme $C$, which is plotted schematically in Fig. \ref{fig:schemes_TAT}(C). We note that scheme $C$ also describes a possibly more realistic scenario where the measured magnetic field cannot be switched off during the state preparation stage of the protocol.

First, the TAT and the magnetic field are applied simultaneously for a time $t'$, so that the initial state evolves by the unitary transformation $\hat{U}_{\omega,\eta}(t') \equiv \exp [ - i t' (\hat{H}_{\omega} + \hat{H}_{\eta})/\hbar]$, where $\hat{H}_\omega + \hat{H}_\eta =  \hbar\omega\hat{J}_y/\sqrt{N} +  i\hbar\eta (\hat{J}_{-}^2 - \hat{J}_{+}^2)/N$ is the sum of the TAT Hamiltonian and the magnetic field Hamiltonian. Following this, we switch off the TAT Hamiltonian and allow the spins to evolve in the magnetic field for a time $t$, resulting in an evolution operator $\hat{D}_\omega (t)$. The final state is thus: \begin{equation} \ket{\psi_C} = \hat{D}_\omega (t) \hat{U}_{\omega,\eta}(t') \ket{\downarrow}^{\otimes N} . \label{eq:psi_C} \end{equation} Again, to ensure that the total time is limited to $\tau$, we have $t' = \tau - t$. Also, if $t=\tau$, there is no two-axis twisting and scheme $C$ reduces to scheme $A$.

As in scheme $B$, the analytic calculation of the quantum Fisher information $F_C$ is intractable, so we calculate it numerically. The solid green lines in Fig. \ref{fig:sens_vs_t_TAT} show the dependence of $\sqrt{F_C}/\tau$ on the sensing time $t/\tau$. We see that scheme $C$ can give better sensitivity than scheme $A$, even in parameter regimes where scheme $B$ gives no advantage over scheme $A$ [see Figs. \ref{fig:sens_vs_t_TAT}(a), \ref{fig:sens_vs_t_TAT}(b) and \ref{fig:sens_vs_t_TAT}(c)]. In such cases, applying the two-axis twisting and the magnetic field simultaneously is a more effective use of the limited time resource $\tau$ then applying them separately (as in scheme $B$) or without any twisting at all (as in scheme $A$).

We can numerically optimise the sensitivity $\sqrt{F_C}/\tau$ over the sensing time $t/\tau$. This is plotted in the solid green lines in Figs. \ref{fig:opt_sens_vs_tau_TAT}(a) and \ref{fig:opt_sens_vs_tau_TAT}(b), with the corresponding optimal sensing times $(t/\tau)_\text{opt}$ plotted in Figs. \ref{fig:opt_sens_vs_tau_TAT}(d) and \ref{fig:opt_sens_vs_tau_TAT}(e), respectively. It appears that scheme $C$ outperforms schemes $A$ and $B$ for all values of $\eta\tau$, with the sensitivities of all three schemes converging to $\sqrt{F}/\tau \to 1$ as $\eta\tau \to 0$. Also, we see that for small values of $\eta\tau$, the optimal sensing time for scheme $C$ is $(t/\tau)_\text{opt} = 0$, i.e., the two-axis twisting and the magnetic field should be applied simultaneously throughout the protocol. This indicates that, contrary to scheme $B$, the twisting dynamics in scheme $C$ plays a positive role for all possible values of the total time $\tau$, the twisting strength $\eta$, and number of spins $N>1$.


As for scheme $B$, it is possible to calculate an analytic expression for the quantum Fisher information $\sqrt{F_C}/\tau$ in the $N\to\infty$ limit. We find (see Appendix for details) that: \begin{equation} \sqrt{F_{C}}/\tau \stackrel{N\to\infty}{\longrightarrow} \left( \frac{t}{\tau} + \frac{1}{2\eta\tau} \right) e^{2\eta\tau (1 - t/\tau)} - \frac{1}{2\eta\tau} . \label{eq:F_C_N_inf} \end{equation} Optimising over the sensing time $t/\tau$ gives: \begin{eqnarray} && \max_{t/\tau}(\sqrt{F_{C}}/\tau) \stackrel{N\to\infty}{\longrightarrow} \frac{1}{2\eta\tau} \left( e^{2\eta\tau} - 1 \right), \quad \\ && (t/\tau)_\text{opt} \stackrel{N\to\infty}{\longrightarrow} 0 , \label{eq:t_opt_C_bosonic} \end{eqnarray} as plotted in Figs. \ref{fig:opt_sens_vs_tau_TAT}(c) and \ref{fig:opt_sens_vs_tau_TAT}(f), respectively. Calculating the ratio \begin{eqnarray} \frac{\max_{t/\tau}(\sqrt{F_{C}}/\tau)}{\max_{t/\tau}(\sqrt{F_{B}}/\tau)} &\stackrel{N\to\infty}{\longrightarrow}& \left\{ \begin{array}{cc} e\left( 1 - e^{-2\eta\tau} \right) & \text{ if } \eta\tau > 0.5 \\ \frac{1}{2\eta\tau} \left( e^{2\eta\tau} - 1 \right) & \text{ if } \eta\tau \leq 0.5 \end{array} \right\} \nonumber \\ &\geq& 1 , \end{eqnarray} shows that, in the $N\to\infty$ limit, scheme $C$ performs just as well as, or outperforms, scheme $B$ for all values of $\eta\tau$. Here, the largest enhancement \begin{equation} \frac{\max_{t/\tau}(\sqrt{F_{C}}/\tau)}{\max_{t/\tau}(\sqrt{F_{B}}/\tau)} \stackrel{N\to\infty}{\longrightarrow} e \approx 2.7 , \end{equation} is achieved as $\eta\tau \to \infty$. Interestingly, from Eq. \ref{eq:t_opt_C_bosonic} we also see that for all values of $\eta\tau$ the optimal strategy is to have the twisting and the magnetic field operating simultaneously throughout the protocol which again, is consistent with our results for finite $N$.

\section{Magnetic field sensing and One-Axis Twisting \label{sec:OAT}}

\begin{figure}
	\includegraphics[width=1\columnwidth]{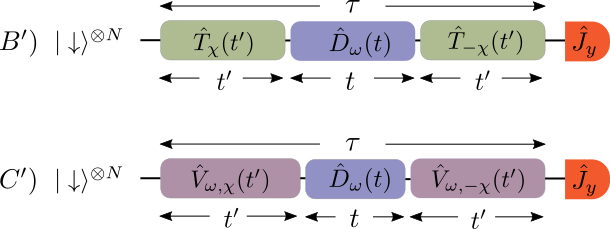}
	\caption{In scheme $B'$, the one-axis twisting operation $\hat{T}_\chi (t')$ generates a spin squeezed state before exposure to the magnetic field through $\hat{D}_\omega (t)$. The ``echo'' (anti-squeezing) operation $\hat{T}_\chi^\dagger (t') = \hat{T}_{-\chi} (t')$ is applied before the final measurement. In scheme $C'$ the spins are exposed to the magnetic field during the OAT and echo operations. For a fair comparison, each protocol is constrained by the time $\tau$.}
\label{fig:schemes_OAT}
\end{figure}

In the previous section we have illustrated the importance of taking state preparation times into account with the example of two-axis twisting. In practice, however, two-axis twisting is difficult to generate. Also, the optimal measurement that was assumed at the readout stage may be difficult to implement in practice, particularly for states that are over-squeezed. In this section we consider two new schemes $B'$ and $C'$ (illustrated in Fig. \ref{fig:schemes_OAT}), which are modifications of schemes $B$ and $C$ of the previous section and are likely to be more feasible in practice.

The new schemes employ one-axis twisting (OAT) instead of two-axis twisting (TAT) in the state preparation stage \cite{ma2011quantum}. OAT has been implemented experimentally in cold atoms \cite{schleier2010squeezing}, atomic vapor-cells \cite{fernholz2008spin} and Bose-Einstein condensates \cite{sorensen2000many,orzel2001squeezed}, for example. For readout, motivated by the recent work of Davis and co-workers \cite{davis2016approaching}, we use an ``echo'' readout protocol. In general, an echo readout applies the inverse of the state preparation operation after the sensing stage, in order to simplify the final measurement \cite{Mac-16} and to overcome strict requirements on the resolution of the final measurement \cite{davis2016approaching, Hos-16}. Such measurements have been implemented in several recent experiments \cite{Lin-16, Hos-16}. However, going beyond previous studies of echo measurements in quantum metrology, we investigate the tradeoffs in sensitivity when a limited time resource must be divided between non-negligible state preparation and readout times and the sensing.

\subsection*{Scheme $B'$}

\begin{figure*}
	\includegraphics[width=2\columnwidth]{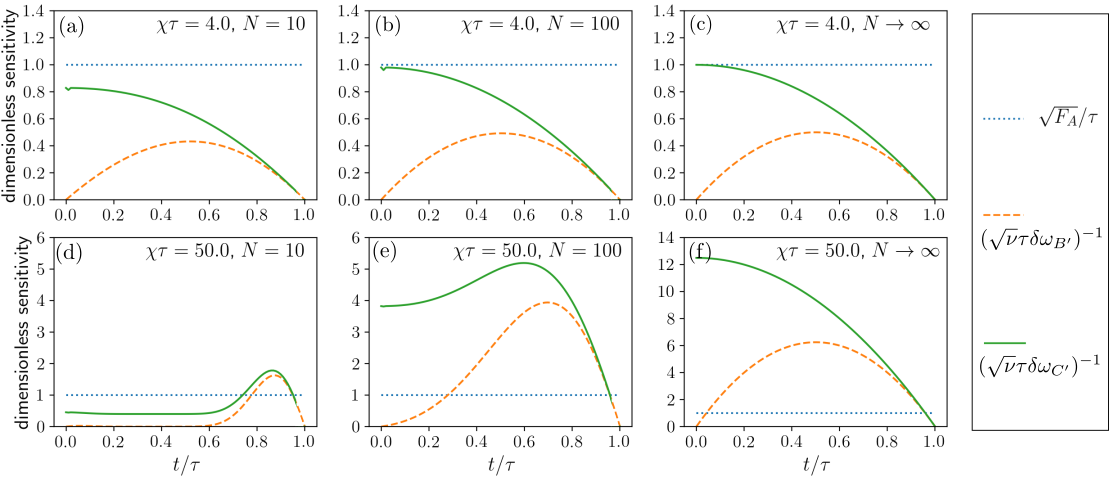}
	\caption{These plots show that for a sufficiently small value of $\chi\tau$ (e.g. $\chi\tau = 4$ and $N=10$), scheme $A$ gives a better sensitivity than scheme $B'$ and scheme $C'$. For a sufficiently large value of $\chi\tau$ (e.g. $\chi\tau = 50$ and $N=10$ or $N=100$), both scheme $B'$ and scheme $C'$ give a better sensitvity than scheme $A$, i.e., the spin squeezing is worthwhile.}
	\label{fig:sens_vs_t_OAT}
\end{figure*}

In our scheme $B'$, starting from the initial state $\ket{\downarrow}^{\otimes N}$, the spins are squeezed by the one-axis twisting (OAT) operation $\hat{T}_{\chi}(t') \equiv \exp [ -i t' \hat{H}_{\chi}/\hbar ]$, where $\hat{H}_\chi = \hbar\chi \hat{J}_x^2 / N$ is the OAT Hamiltonian, $\chi$ is the spin squeezing strength, and $t'$ is the state preparation time. Similar to TAT, OAT generates spin squeezed states for short state preparation times and over-squeezed states (such as Schr\"{o}dinger cat states) for longer state preparation times. After the spins are prepared in the twisted state, they are exposed to the magnetic field for a time $t$, resulting in a rotation of the state around the spin $y$-axis by $\hat{D}_\omega (t)=\exp [ - i t\hat{H}_{\omega}/\hbar ]$. For readout, we use an echo measurement. An echo measurement applies the inverse of the state preparation operation after the sensing stage, in order to simplify the final measurement. Since, in our case, the state preparation is the OAT operation $\hat{T}_{\chi}(t')$, we apply the inverse operation $\hat{T}_\chi^\dagger (t') = \hat{T}_{-\chi} (t')$, after the sensing stage. The final state is thus: \begin{equation} \ket{\psi_{B'}} = \hat{T}_{-\chi} \left( t' \right) \hat{D}_\omega (t) \hat{T}_\chi \left( t' \right) \ket{\downarrow}^{\otimes N} . \label{eq:psi_B'} \end{equation} To ensure that the total time of scheme $B'$ is limited to $\tau$, we have $t' = (\tau - t)/2$. Finally, after the echo, we measure the collective observable $\hat{J}_y$. By the propagation of error formula, the error in the estimate of the small scaled magnetic field $\omega$ is:
\begin{align}
 \delta\omega_{B'} = \frac{1}{\sqrt{\nu}} \left| \frac{\Delta \hat{J}_y}{ \partial_\omega \braket{\hat{J}_y} } \right|_{\omega = 0} ,
\label{sens_B'}
\end{align}
where $| \Delta \hat{J}_y |_{\omega = 0} = \sqrt{N}/2$ is the standard deviation of the measured operator $\hat{J}_y$ in the state $\ket{\psi_{B'}}$, and (see Appendix for details of the calculation):
\begin{align}
 \left| \partial_\omega \braket{\hat{J}_y} \right|_{\omega = 0} = \frac{t \sqrt{N}(N-1)}{2} \left| \sin\theta(t) \cos^{N-2}\theta(t) \right| ,
\label{eq:d_w_EJy}
\end{align}
where $\braket{\hat{J}_y}$ is the expectation value and $\theta(t) = \chi\tau(1 - t/\tau) / (2N)$. Substituting into Eq. \ref{sens_B'} gives an expression for $\delta\omega_{B'}$, which in turn can be used to calculate the dimensionless sensitivity
%

\begin{align}
 (\sqrt{\nu}\tau\delta\omega_{B'})^{-1} = (t/\tau) (N-1) \left| \sin\theta(t) \cos^{N-2}\theta(t) \right| . \label{eq:w_B'}
\end{align}
From Eq. \ref{eq:w_B'} it is straightforward to see that $(\sqrt{\nu} \tau \delta\omega_{B'})^{-1}$ depends on only the three dimensionless variables $N$, $t/\tau$, and $\chi\tau$. In Fig. \ref{fig:sens_vs_t_OAT}, we plot $(\sqrt{\nu}\tau\delta\omega_{B'})^{-1}$ as a function of the sensing time $t/\tau$ for various choices of $\chi\tau$ and $N$ (the dashed orange lines). We see that, depending on the values of $\chi\tau$ and $N$, there is a $t/\tau$ that optimises the sensitivity. This optimisation is done numerically and the results are plotted against $\chi\tau$ in Fig. \ref{fig:opt_sens_vs_tau_OAT} (the dashed orange line) showing that scheme $B'$ behaves in a similar fashion to the analagous TAT scheme $B$ in that it is not always guaranteed to give sensitivity gains relative to scheme $A$. For example, for $N = 10$ we must have $\chi \tau \gtrsim 11.5$, for scheme $B'$ (the dashed orange line) to outperform scheme $A'$ (the dotted blue line). This indicates that if the total measurement time $\tau$ is too short, or the squeezing strength $\chi$ is too weak, the limited time resource is used more effectively by devoting more time to probing the magnetic field and less time to spin squeezing. For $N = 100$ the threshold for scheme $B'$ to outperform scheme $A$ is $\chi \tau \gtrsim 8.2$, a lower value than for $N=10$ [see Fig. \ref{fig:opt_sens_vs_tau_OAT}(b)]. This suggests that as $N$ increases, it becomes possible to beat scheme $A'$ with a shorter total time $\tau$ (or weaker squeezing strength $\chi$). Below we will see that as $N\to\infty$ this threshold value saturates at $\chi\tau > 8$.

To find the sensitivity for scheme $B'$ in the $N\to\infty$ limit, we can simply take the $N\to\infty$ limit of Eq. \ref{eq:w_B'}. We find: \begin{equation} (\sqrt{\nu}\tau\delta\omega_{B'})^{-1} \stackrel{N\to\infty}{\longrightarrow} \frac{\chi t(\tau - t)}{2\tau} . \label{eq:sens_B'_bosonic} \end{equation} Unlike the finite-$N$ case where numerical optimisation was necessary, we can easily optimise Eq. \ref{eq:sens_B'_bosonic} to find the analytic expression: \begin{equation} \max_{t/\tau} (\sqrt{\nu}\tau\delta\omega_{B'})^{-1}  \stackrel{N\to\infty}{\longrightarrow} \frac{\chi\tau}{8}, \quad (t/\tau)_\text{opt} \stackrel{N\to\infty}{\longrightarrow} \frac{1}{2} . \end{equation} Comparison with the sensitivity $(\sqrt{\nu}\tau\delta\omega_{A})^{-1} = 1$ for scheme $A$ shows that preparation of a twisted state via scheme $B'$ is worthwhile only if $\chi\tau > 8$. If $\chi\tau < 8$, however, scheme $B'$ gives a worse sensitivity than scheme $A$, since the time cost of preparing the twisted state outweighs any benefits of twisting. Since $(t/\tau)_\text{opt} \stackrel{N\to\infty}{\longrightarrow} 1/2$ we also conclude that scheme $B'$ is optimised by using half of the total available time $\tau$ for sensing, and a quarter each, $(t'/\tau)_\text{opt} \stackrel{N\to\infty}{\longrightarrow} 1/4$, for preparation of the squeezed state and the echo readout.

\subsection*{Scheme $C'$}

In analogy with scheme $C$ for TAT, in scheme $C'$ we suppose that during the state preparation stage the OAT and the magnetic field are applied simultaneously, so that the initial state evolves by the unitary transformation $\hat{V}_{\omega,\chi}(t') \equiv \exp [ - i t' (\hat{H}_{\omega} + \hat{H}_{\chi})/\hbar]$, where $\hat{H}_\omega + \hat{H}_\chi = \hbar\omega \hat{J}_y /\sqrt{N} + \hbar\chi \hat{J}_x^2/N$. Following this, we switch off the OAT and allow the spins to evolve in the magnetic field for a time $t$, resulting in an evolution operator $\hat{D}_\omega (t)$. Finally, we implement the echo readout by reversing the one-axis twisting component (but not the magnetic field component) of the state preparation with the operation $\hat{V}_{\omega,-\chi}(t')$. The final state is thus: \begin{equation} \ket{\psi_{C'}} = \hat{V}_{\omega, -\chi} (t') \hat{D}_\omega (t) \hat{V}_{\omega,\chi}(t') \ket{\downarrow}^{\otimes N} . \label{eq:psi_C'} \end{equation} Again, to ensure that the total time is limited to $\tau$, we have $t' = (\tau - t)/2$. As in scheme $B'$, we measure the collective observable $\hat{J}_y$, giving the error:

\begin{align}
 \delta\omega_{C'} = \frac{1}{\sqrt{\nu}} \left| \frac{\Delta \hat{J}_y}{ \partial_\omega \braket{\hat{J}_y} } \right|_{\omega = 0} .
\label{sens_C'}
\end{align}

\begin{figure*}
	\includegraphics[width=2\columnwidth]{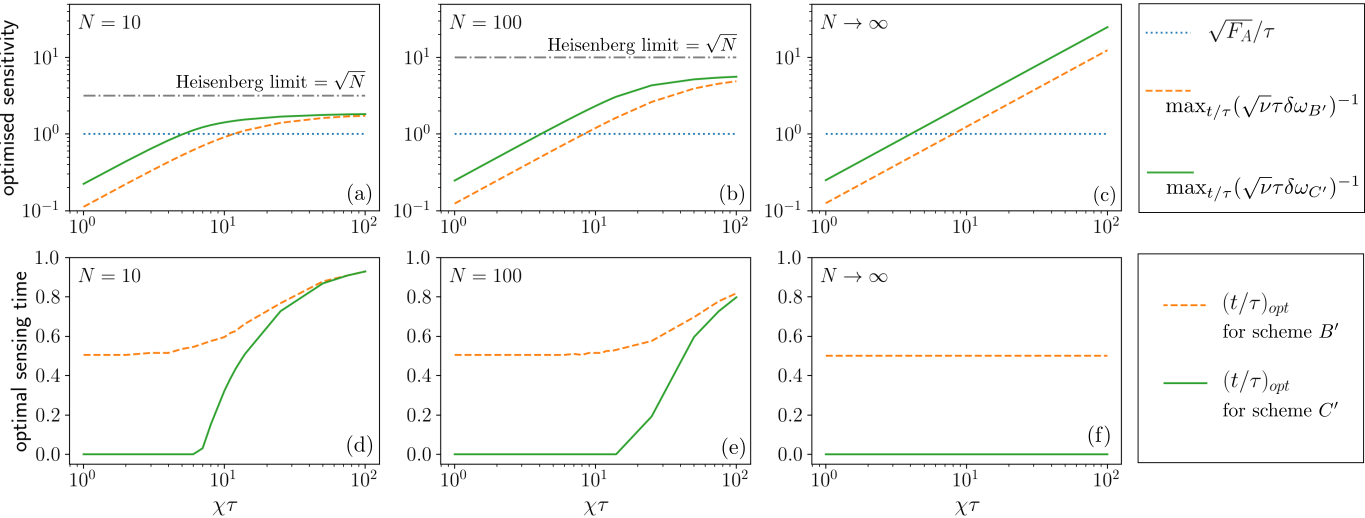}
	\caption{The upper plots show the optimised sensitivity $\max_{t/\tau}(\sqrt{\nu}\tau\delta\omega)^{-1}$ as a function of $\chi\tau$. These plots are optimised over time but not over measurements in contrast to Fig.\ref{fig:sens_vs_t_TAT} and Fig.\ref{fig:opt_sens_vs_tau_TAT} which are optimised over both. Comparison of schemes reveals that when $N=10$ scheme $A$ outperforms scheme $B$ for $\chi\tau\lesssim 11.5$ and scheme C for $\chi\tau \lesssim 5$. These threshold values decrease for larger $N$. For very large $\chi\tau$, the sensitivities of schemes $B$ and $C$ converge. The lower plots show the optimal sensing time $(t/\tau)_\text{opt}$ as a function of $\chi\tau$.}
	\label{fig:opt_sens_vs_tau_OAT}
      \end{figure*}

At $\omega = 0$ the standard deviation in the numerator is just that of the initial state, $| \Delta \hat{J}_y |_{\omega = 0} = \sqrt{N}/2$. However, the denominator cannot be easily calculated analytically, so we pursue a numerical approach. Fig. \ref{fig:sens_vs_t_OAT} shows the dependence of the dimensionless sensitivity $(\sqrt{\nu}\tau\delta\omega_{C'})^{-1}$ on the sensing time $t/\tau$ for scheme $C'$ (the solid green lines). After optimising over the sensing time $t/\tau$, as depicted in Fig. \ref{fig:opt_sens_vs_tau_OAT}, it becomes apparent that scheme $C'$ (the solid green line) always outperforms scheme $B'$. Additionally, it is also clear from Fig. \ref{fig:opt_sens_vs_tau_OAT}, that scheme $C'$ gives an advantage over scheme $A$ for a wider range of values of $\chi\tau$ than does scheme $B'$. When $N=10$, for example, scheme $C'$ beats scheme $A$ if $\chi\tau \gtrsim 5$, compared to $\chi\tau \gtrsim 11.5$ for scheme $B'$ [see Fig. \ref{fig:opt_sens_vs_tau_OAT}(a)].  

We now analyse scheme $C'$ in the $N\to\infty$ limit. For finite-$N$, due to the difficulty of analytic calculation we found the sensitivity numerically (as shown in the solid green lines of Figs. \ref{fig:sens_vs_t_OAT} and \ref{fig:opt_sens_vs_tau_OAT}). However, in the $N\to\infty$ limit it is possible to derive the analytic expression (see Appendix for details): \begin{equation} (\sqrt{\nu}\tau \delta\omega_{C'})^{-1} \stackrel{N\to\infty}{\longrightarrow} \frac{\chi\tau}{4} \left( 1 - t^2 / \tau^2 \right) . \label{eq:sens_C'} \end{equation} Optimising over the sensing time $t/\tau$ gives: \begin{equation} \max_{t/\tau} (\sqrt{\nu}\tau\delta\omega_{\tilde{C}})^{-1} \stackrel{N\to\infty}{\longrightarrow} \frac{\chi\tau}{4}, \quad (t/\tau)_\text{opt} \stackrel{N\to\infty}{\longrightarrow} 0 , \end{equation} a factor of 2 improvement on the sensitivity over the corresponding $N\to\infty$ version of scheme $B'$. Squeezing via scheme $\tilde{C}$ gives a better sensitivity than scheme $A$ provided that $\chi\tau > 4$, but a worse sensitivity if $\chi\tau < 4$. Also, we note that in agreement with the $N\to\infty$ limit of the TAT scheme $C$ in the previous section, the optimal sensitivity for scheme $C'$ is achieved for $(t/\tau)_\text{opt} \stackrel{N\to\infty}{\longrightarrow} 0$, so that the twisting and magnetic field should both be operating at all times in the protocol.

\section{Conclusions\label{sec8}}

It is a well known result in the field of quantum metrology that preparation of an entangled probe state before sensing can, in principle, give a factor of $\sqrt{N}$ enhancement over the optimal sensitivity with separables states. However, it is usually assumed that state preparation and readout times are negligible. In this paper we have shown that when the total available time $\tau$ is a limited resource, entangled state preparation is not always worthwhile when non-negligible state preparation and readout times are taken into account. In particular, for magnetic field sensing with twisted states, it is more advantageous to devote all of the available time to sensing if the twisting strength is sufficiently weak, or the total available time sufficiently short. However, in the case where the twisting is strong enough that entangled state preparation is worthwhile, we have also shown that a more effective use of time is to `blend' the state preparation, sensing and readout stages by allowing the twisting dynamics and the magnetic field to operate concurrently. This also corresponds to the (possibly more realistic) scenario where the magnetic field cannot be switched off during the state preparation and readout.

By a combination of analytics and numerics, our results cover a broad range of parameters, from small $N$ to $N\to\infty$. We note that by the Holstein-Primakoff transformation \cite{holstein1940field} (see Appendix), there is an exact correspondence between a spin system in the $N\to\infty$, and a bosonic field mode (the ``bosonic limit''). This extends our results into a setting where, instead of magnetic field sensing with a twisted state of spins, we are sensing the displacement of a bosonic field mode with squeezed states.

We note that an important assumption in this paper is that the total available time $\tau$ is a limited resource. In practice this limit could be enforced, for example, by decoherence, by the stability of our equipment or by the fact that the quantity we want to measure is rapidly changing. Future work could include the effects of decoherence in the state preparation, sensing and readout stages. Further work could also include investigation into the experiment demonstrated by M. Penasa \textit{et.al} \cite{penasa2016measurement} in which an echo measurement protocol is employed to estimate the amplitude of a small displacement acting on a cavity field. The notable difference in the scheme of Penasa and the schemes analysed here is that execution of preparation and readout takes the form of atom-cavity interactions in order to create, and undo the creation of, optical cat states.



%
\section*{Acknowledgments}
The authors thank Y. Matsuzaki for many useful discussions.\\

The authors acknowledge support from the United Kingdom EPSRC through the Quantum Technology Hub: Networked Quantum Information Technology (grant reference EP/M013243/1) and the MEXT KAKENHI Grant-in-Aid for Scientific Research on Innovative Areas Science of hybrid quantum systems Grant No.15H05870. 
\newpage
\bibliography{Sim}
\bibliographystyle{ieeetr}

\appendix
\section*{The Bosonic ($N\to\infty$) limit}
The Holstein-Primakoff transformations \cite{holstein1940field} allow us to map the $N$-spin system in the $N\to\infty$ limit to a bosonic field mode. We have:
\begin{eqnarray}
\tilde{a} &=& \lim_{N\to\infty}\frac{\hat{J}_{-}}{\sqrt{N}}, \label{HP_1} \\
\tilde{a}^\dagger &=& \lim_{N\to\infty}\frac{\hat{J}_{+}}{\sqrt{N}}, \label{HP_2} \\
\ket{0} &=& \lim_{N\to\infty} \ket{\downarrow}^{\otimes N}, \label{HP_3}
\end{eqnarray} where $\tilde{a},\:\tilde{a}^{\dagger}$ are the bosonic annihilation operators which obey the bosonic commutation relation $[\tilde{a},\tilde{a}^{\dagger}]=1$, and $\ket{0}$ is the bosonic vacuum state. By taking the $N\to\infty$ limit of all operators and states in schemes $A$, $B$, $C$, $B'$ and $C'$ we can thus use Eqs. \ref{HP_1}--\ref{HP_3} find the corresponding operators for sensing schemes with a bosonic mode as the probe system. For instance the spin rotation operator $\hat{D}_\omega (\tau)$ becomes, after the $N\to\infty$ limit, the bosonic displacement operator \begin{equation} \tilde{D}_\omega (\tau) \equiv \lim_{N\to\infty} \hat{D}_\omega (\tau) = e^{\tau\omega (\tilde{a} - \tilde{a}^\dagger) / 2} , \end{equation} where, to avoid confusion with spin operators, a tilde above an operator denotes a bosonic mode operator and, again, $\omega$ is the parameter to be estimated. Here, the parameter $\omega$ could be, for example, a weak classical force acting on an harmonic oscillator \cite{munro2002weak,braginsky1995quantum}, or an electric field applied to an optical field mode in a cavity. Similarly, the $N\to\infty$ limit of the TAT operator is the bosonic quadrature squeezing operator \begin{eqnarray} \tilde{S}_\eta (t') &\equiv& \lim_{N\to\infty}\hat{S}_\eta (t') = e^{t'\eta (\tilde{a}^2 - \tilde{a}^{\dagger 2})} . \end{eqnarray} Also, \begin{eqnarray} \tilde{U}_{\omega,\eta} (t') &\equiv& \lim_{N\to\infty}\hat{U}_{\omega,\eta} (t') = e^{t'\omega (\tilde{a} - \tilde{a}^\dagger) / 2 + t'\eta (\tilde{a}^2 - \tilde{a}^{\dagger 2})} . \quad \end{eqnarray} These squeezing operations $\tilde{S}_\eta$ and $\tilde{U}_{\omega,\eta}$ can be implemented in optical systems, for example, via parametric down conversion in nonlinear crystals \cite{gerry2005introductory,lvovsky2014squeezed,boyd2003nonlinear}.

\section*{Deriving $\sqrt{F_B}/\tau$ in the $N\to\infty$ limit (Eq. \ref{eq:F_B_N_inf})}

Applying the definition of the quantum Fisher information, Eq. \ref{eq:QFI}, to the state \begin{equation} \lim_{N\to\infty}\ket{\psi_B} = \tilde{D}_\omega(t) \tilde{S}_\eta(t') \ket{0} , \end{equation} and making use of the identity $\tilde{S}_\eta^\dagger(t') \tilde{a} \tilde{S}_\eta (t') = \tilde{a} \cosh (2\eta t') - \tilde{a}^\dagger \sinh (2\eta t')$ \cite{gerry2005introductory}, we find that \begin{equation} (\sqrt{\nu}\tau\delta\omega_{\tilde{B}})^{-1} \leq \sqrt{F_{\tilde{B}}}/\tau = \frac{t}{\tau} e^{2\eta\tau (1 - t/\tau)} . \label{eq:sens_B_bosonic} \end{equation} 

\section*{Deriving $\sqrt{F_C}/\tau$ in the $N\to\infty$ limit (Eq. \ref{eq:F_C_N_inf})}

Applying the definition of the quantum Fisher information in Eq. \ref{eq:QFI} to the state \begin{equation} \lim_{N\to\infty} \ket{\psi_C} = \tilde{D}_\omega (t) \tilde{U}_{\omega,\eta}(t')\ket{0} , \end{equation} and using the expansion $e^{X}Ye^{-X} = Y + [X,Y] + \frac{1}{2!}[X,[X,Y]] + ...$, we find that \begin{equation} (\sqrt{\nu}\tau\delta\omega_{\tilde{C}})^{-1} \leq \sqrt{F_{\tilde{C}}}/\tau = \left( \frac{t}{\tau} + \frac{1}{2\eta\tau} \right) e^{2\eta\tau (1 - t/\tau)} - \frac{1}{2\eta\tau} . \end{equation}  

\section*{Deriving Eq. \ref{eq:d_w_EJy}}

Here we follow the derivation given in Ref. \cite{davis2016approaching}. Using the expression for $\ket{\psi_{B'}}$ in Eq. \ref{eq:psi_B'}, one can show that \begin{equation} | \partial_\omega \langle \hat{J}_y \rangle |_{\omega = 0} = \left| \frac{it}{\sqrt{N}} \bra{\downarrow}^{\otimes N} \left[ \hat{T}_{-\chi}(t')\hat{J}_y \hat{T}_{\chi}(t') , \hat{J}_y \right] \ket{\downarrow}^{\otimes N} \right| , \end{equation} where $\langle \hat{J}_y \rangle = \bra{\psi_{B'}} \hat{J}_y \ket{\psi_{B'}}$. The operator in the commutator can be expressed as:

\begin{eqnarray} \hat{T}_{-\chi}(t')\hat{J}_y \hat{T}_{\chi}(t') &=& e^{it'\chi \hat{J}_x^2 /N} \left(-\frac{i}{2}\hat{J}_+ + \frac{i}{2} \hat{J}_{-} \right) e^{-it'\chi \hat{J}_x^2 /N} \nonumber \\ &=& e^{-i\pi\hat{J}_y/2} \Big[ -\frac{i}{2} e^{it'\chi (2\hat{J}_z -1)/N} \hat{J}_+ + \nonumber\\ &&  + \frac{i}{2} \hat{J}_{-} e^{it'\chi (-2\hat{J}_z -1)/N} \Big] e^{i\pi\hat{J}_y/2} . \end{eqnarray} Substituting back into the expression for $| \partial_\omega \langle \hat{J}_y \rangle |_{\omega = 0}$ above gives a long expression containing expectation values of the sort: \begin{equation} \bra{+}^{\otimes N} \hat{J}_{-}^2 e^{-2it' \chi \hat{J}_z} \ket{+}^{\otimes N} , \label{eq:blah} \end{equation} for example. Such expectation values can be calculated by differentiating the generating function given in the appendix of Ref. \cite{arecchi1972atomic}. For example, in \cite{arecchi1972atomic} it was shown that: \begin{eqnarray} X_A(\alpha,\beta,\gamma) &\equiv& \bra{+}^{\otimes N} e^{\gamma \hat{J}_-} e^{\beta \hat{J}_z} e^{\alpha \hat{J}_+}  \ket{+}^{\otimes N} \\ &=& \left[ \frac{1}{2} e^{-\beta/2} + \frac{1}{2} e^{\beta/2}(\alpha+1)(\gamma + 1) \right]^N \end{eqnarray} Eq. \ref{eq:blah} is then calculated as: \begin{eqnarray} && \bra{+}^{\otimes N} \hat{J}_{-}^2 e^{-2it' \chi \hat{J}_z} \ket{+}^{\otimes N} = \left[ \frac{\partial^2}{\partial\gamma^2} X_A \right]_{\substack{ \alpha = \gamma = 0 \\ \beta = -2it'\chi / N}} \\ && \qquad = \frac{N(N-1)}{4} \left( \cos\frac{t'\chi}{N} \right)^{N-2} e^{-2it'\chi /N} . \nonumber \end{eqnarray} Using this procedure on all terms leads to the final expression:

\begin{equation} | \partial_\omega \langle \hat{J}_y \rangle |_{\omega = 0} = \frac{t \sqrt{N}(N-1)}{2} \left| \sin\theta(t) \cos^{N-2}\theta(t) \right| . \end{equation}

\section*{Deriving Eq. \ref{eq:sens_C'}}

In the $N\to\infty$ limit the final state at the end of scheme $C'$ is: \begin{equation} \lim_{N\to\infty} \ket{\psi_{C'}} = \tilde{V}_{\omega, -\chi}(t') \tilde{D}_\omega (t) \tilde{V}_{\omega,\chi} (t') \ket{0} , \end{equation} where \begin{eqnarray} \tilde{V}_{\omega,\chi} (t') &\equiv& \lim_{N\to\infty}\hat{V}_{\omega,\chi}(t') \\ &=& \exp\left[ t' \omega (\tilde{a} - \tilde{a}^\dagger) - it' \chi (\tilde{a} + \tilde{a}^\dagger)^2 \right] \end{eqnarray} is found by applyication of the Holstein-Primakoff transformations. Defining $\tilde{P} = -i\tilde{a}^\dagger + i\tilde{a}$, it is straightforward to calculate $\left| \Delta \tilde{P} \right|_{\omega = 0} = 1$, where $\Delta \tilde{P}$ is the standard deviation of $\tilde{P}$ in the state $\lim_{N\to\infty}\ket{\psi_{C'}}$. Next, by repeated use of the expansion $e^{X}Ye^{-X} = Y + [X,Y] + \frac{1}{2!}[X,[X,Y]] + ...$, we find that the expectation value of $\tilde{P}$ is $\langle\tilde{P}\rangle = \frac{1}{4} (\tau^2 - t^2)\chi\omega$. Now, since \begin{equation} \lim_{N\to\infty}\delta\omega_{C'} = \frac{1}{\sqrt{\nu}} \left| \frac{\Delta \tilde{P}}{\partial_\omega \langle \tilde{P} \rangle} \right|_{\omega = 0} , \end{equation} we can substitute the expressions above to find: \begin{equation} (\sqrt{\nu}\tau \delta\omega_{\tilde{C}})^{-1} = \frac{\chi\tau}{4} \left( 1 - t^2 / \tau^2 \right) . \end{equation}

\end{document}